# Thermal Characterization of Buried Interfaces in Multilayer Heterostructures via TDTR with Periodic Waveform Analysis

Mingzhen Zhang[1], Puqing Jiang[1,*], Ronggui Yang[1,2,*]

[1]School of Energy and Power Engineering, Huazhong University of Science and Technology, Wuhan, Hubei 430074, China

[2]School of Mechanics and Engineering Science, Peking University, Beijing 100871, China

**Abstract**

Accurate evaluation of buried thermal interfaces is vital for understanding and optimizing heat dissipation in wide- and ultra-wide-bandgap (WBG/UWBG) semiconductor devices. Conventional time-domain thermoreflectance (TDTR) typically probes only near-surface transport due to its restricted modulation frequency range. Here, we employ a frequency-tunable periodic waveform analysis TDTR (PWA-TDTR) technique to perform depth-resolved thermal measurements on three representative systems: epitaxial ε-$Ga_2O_3$/SiC, GaN/Si, and mechanically bonded GaN/diamond. By combining broadband multi-frequency probing with sensitivity-guided joint fitting, we quantitively determine interfacial thermal conductance, layer-specific thermal conductivity, and volumetric heat capacity, without requiring destructive sample preparation. The results reveal that the buried $Ga_2O_3$/SiC interface exhibits weak phonon transmission due to acoustic mismatch; the transition layers in GaN/Si act as phonon-impedance gradients that redistribute heat flux; and the GaN/diamond boundary remains the dominant thermal bottleneck despite diamond's ultrahigh bulk conductivity. These findings demonstrate that the modulation frequency in PWA-TDTR functions as a tunable probe of depth-dependent phonon transport, directly linking frequency-domain thermal response to interfacial energy transmission. Overall, this work positions PWA-TDTR as a versatile platform for investigating buried nonmetal–nonmetal interfaces in next-generation high-power and optoelectronic materials.

*Corresponding Authors: jpq2021@hust.edu.cn (P. Jiang), ronggui@pku.edu.cn (R. Yang)

# 1. Introduction

Efficient thermal management remains a key bottleneck in wide-bandgap (WBG) and ultra-wide-bandgap (UWBG) electronic devices, where self-heating limits reliability, output power, and scaling potential [1–3]. A common strategy to mitigate these issues is heterogeneous integration, which involves bonding active layers to high-thermal-conductivity substrates, such as diamond [4,5] or SiC [6,7], or inserting intermediate buffer layers to spread heat laterally. Yet, in these complex stacks, heat dissipation is often governed not by the bulk materials but by the buried thermal boundary conductance (TBC) at nonmetal–nonmetal interfaces, where phonon coupling is intrinsically poor. Despite their technological importance, quantitative studies of buried TBC remain limited: fewer than 50 interfaces have been systematically characterized [8], and most reported data concern metal–nonmetal interfaces measured by time-domain thermoreflectance (TDTR) [9–11].

Pump–probe thermoreflectance methods such as TDTR and frequency-domain thermoreflectance (FDTR) [12–14] have become indispensable for microscale thermal characterization. By modulating a focused pump laser and detecting reflectance changes either as a function of delay time or modulation frequency, these techniques offer non-contact access to thermal conductivity and TBC. Conventional TDTR, however, typically operates at modulation frequencies of 0.1-10 MHz, giving thermal penetration depths of only a few microns [15]. This limitation restricts its sensitivity to shallow interfaces and makes parameter extraction in multilayer systems nontrivial. FDTR can access a broader frequency range, but unlike time-domain approaches that directly sample the transient response, FDTR's interpretation depends on frequency-domain modeling and accurate calibration of the phase reference and system transfer function, which can complicate robust multiparameter inversion in complex multilayers.

Recent developments have sought to overcome these depth and robustness constraints. Steady-state thermoreflectance (SSTR) achieves low-frequency heating suitable for buried substrates several micrometers thick [16]. Dual-modulation TDTR imaging combines steady-state and MHz transients to map both thermal conductivity and TBC in $\beta$-Ga$_2$O$_3$/SiC heterostructures [17]. Micro-structured confinement in FDTR has been used to enhance thermal-wave penetration and quantify TBC across SiO$_2$/Al$_2$O$_3$ buried interfaces [18]. Interface engineering strategies such as surface-activated bonding have been shown to raise GaN/SiC TBC above 200 MW/(m$^2$·K) [19]. Additional progress includes discrete spectral Bayesian deconvolution for complex multilayer inversion [20], multi-pulse thermoreflectance imaging with structure-function analysis for improved depth sensitivity [21], the three-sensor 3ω–2ω method for simultaneous extraction of film thermal conductivity and interfacial thermal resistance [22], and pulsed thermoreflectance imaging applied to GaN heterostructures [23]. Collectively, these studies underscore the growing demand for high-resolution, depth-resolved thermal metrology in multilayer systems where buried interfaces govern heat dissipation. Yet most methods remain system-specific, require modified optics, or lack broad applicability to structures containing thick intermediate layers or multiple buried interfaces.

To address this gap, we employ periodic waveform analysis-based time-domain thermoreflectance (PWA-TDTR) [24], which extends TDTR measurements to modulation frequencies as low as 50 Hz without altering the optical configuration. By reconstructing the time-dependent thermal response under periodic excitation, PWA-TDTR greatly increases the thermal penetration depth and enables frequency-selective sensitivity to interfaces at different depths. This capability makes it particularly suited for realistic multilayer heterostructures such as Ga$_2$O$_3$-on-SiC, GaN-on-Si, and GaN-on-diamond, all of which contain relatively thick transition regions. In this work, we use PWA-TDTR to perform systematic, depth-resolved measurements on these three representative systems. By combining multi-frequency datasets and with sensitivity-guided joint fitting, we quantitatively determine interfacial thermal conductance together with the thermal conductivity and volumetric heat capacity of individual layers without destructive sample preparation. Beyond demonstrating experimental feasibility, these results provide materials-specific insights into phonon transport across buried interfaces and illustrate how bonding quality and transition-layer design control heat dissipation in next-generation WBG and UWBG devices.

## 2. Experimental Method and Measurement Principle

In this work, we employed the PWA-TDTR technique to achieve depth-resolved and quantitative thermal characterization of multilayer heterostructures. The schematic of the experimental setup is shown in Figure 1. Unlike conventional TDTR, which is primarily sensitive to near-surface transport, the PWA-TDTR configuration integrates a high-performance lock-in amplifier and advanced signal-processing module while preserving the original optical layout and laser source. These improvements enable stable signal acquisition under low-frequency modulation, thereby greatly enhancing sensitivity to heat diffusion through buried interlayers and interfaces—an essential requirement for realistic device structures.

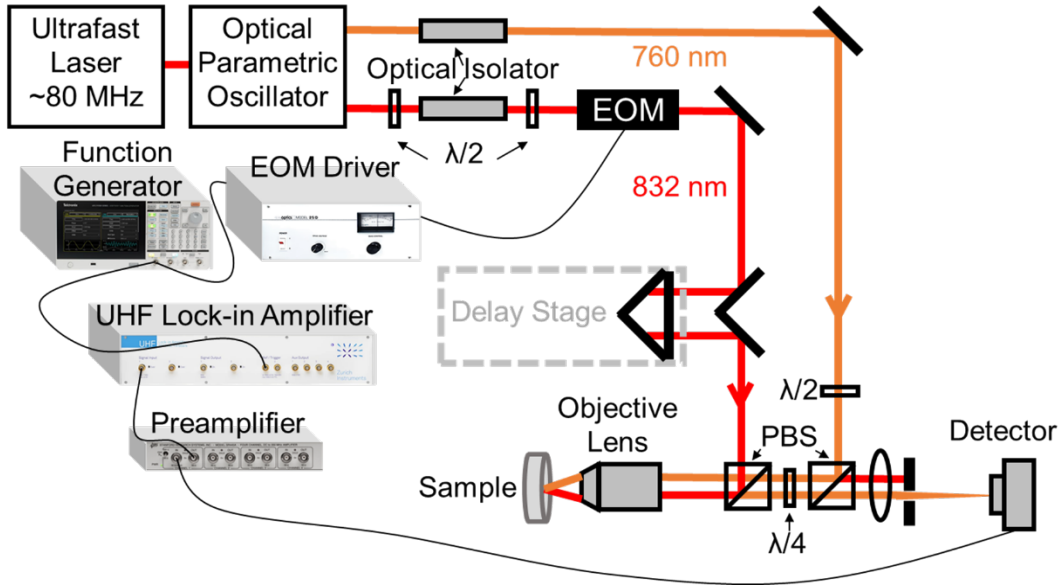

**Fig. 1.** Schematic of the PWA-TDTR setup (PBS: polarizing beam splitter; EOM: electro-optic modulator; λ/2: halfwave plate; λ/4: quarter-wave plate).

The PWA-TDTR system preserves all functional components of a standard TDTR apparatus but incorporates two key upgrades: a balanced photodetector (PDA465A, Thorlabs) for noise suppression and a UHFLI lock-in amplifier (Zurich Instruments) equipped with periodic waveform analysis (PWA) functionality. This advanced lock-in architecture enables simultaneous acquisition of multiple harmonic components and reconstruction of the full periodic temperature waveform at the sample surface, providing richer thermal information compared to the sinusoidal detection scheme of conventional TDTR.

For low-frequency operation, the UHFLI lock-in amplifier generates a 50% duty-cycle square-wave reference to drive the electro-optic modulator (EOM), while synchronously recording the reflected probe signal. In contrast to conventional TDTR, the delay stage is fixed at -20 ps, and the mechanical chopper is removed, allowing PWA functionality to capture the complete thermoreflectance waveform throughout the modulation cycle. The acquired data are normalized in amplitude and time, and subsequently fitted using multilayer heat transport models to extract key thermal parameters, including thermal conductivity, volumetric heat capacity, and interfacial thermal conductance.

Further details of the system configuration, signal modeling, and fitting methodology can be found in our previous publication [24], where the reliability of the approach has been systematically benchmarked.

A distinctive advantage of PWA-TDTR is its ability to extend modulation frequencies down to ~50 Hz without altering the optical layout [24]. This dramatically increases the thermal penetration depth, enabling sensitivity to structures from a few micrometers up to several tens of micrometers in thickness. As such, the method is particularly suitable for characterizing thick transition layers, low-thermal-conductivity interlayers, and buried interfaces that are inaccessible to conventional TDTR. In addition, by employing a multi-frequency joint fitting strategy guided by sensitivity analysis [25], the method achieves stable parameter extraction and effective decoupling of thermal contributions from shallow and deep regions—an essential capability for analyzing realistic multilayer heterostructures in advanced devices.

## 3. Sample Structures and Experimental Design

Buried interfaces in advanced semiconductor heterostructures are often the primary bottlenecks for heat dissipation in high-power and high-frequency devices. The interfacial thermal conductance is strongly influenced by lattice mismatch, defect density, bonding quality, and the presence of intermediate layers, all of which ultimately dictate device performance and reliability [26–29]. In this work, we investigate three representative multilayer systems—$Ga_2O_3$-on-SiC, GaN-on-Si, and mechanically bonded GaN-on-diamond—that capture distinct interfacial configurations and integration strategies. These systems span both epitaxial and mechanically bonded architectures, thereby enabling a comparative evaluation of how intrinsic material properties and fabrication routes govern cross-interface thermal transport.

a) $Ga_2O_3$-on-SiC epitaxial structure: This configuration combines an ultra-wide-bandgap semiconductor ($Ga_2O_3$) with a high-thermal-conductivity substrate (SiC). Despite the excellent thermal properties of SiC, the large phonon mismatch across the epitaxial interface results in a relatively high interfacial thermal resistance, which critically impacts the heat removal capability of $Ga_2O_3$-based power devices.

b) GaN-on-Si epitaxial structure: Widely adopted in high-power and RF electronics due to the low cost and scalability of Si substrates, this system suffers from a substantial lattice and thermal expansion mismatch between GaN and Si. The resulting dislocations, residual stresses, and transition layers significantly degrade interfacial thermal conductance, making the GaN/Si interface one of the dominant thermal bottlenecks in GaN device technology.

c) Mechanically bonded GaN-on-diamond trilayer structure: This system is fabricated by bonding a GaN-on-Si epitaxial wafer to a diamond substrate, thereby replicating the thermal dissipation pathway in state-of-the-art GaN power devices. In such a structure, heat generated in the GaN layer must traverse multiple buried interfaces—including Al/GaN, GaN/Si, and GaN/diamond—before reaching the ultra-high-thermal-conductivity diamond substrate. The quality of mechanical bonding, residual interlayers, and interface roughness strongly influence the overall thermal resistance of the stack.

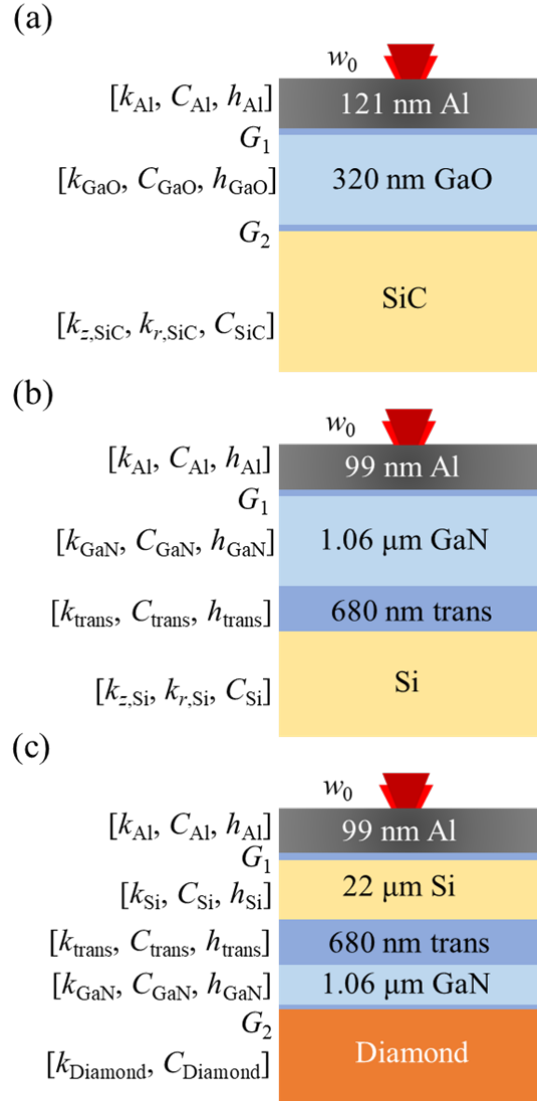

**Fig. 2.** Schematic diagram of the thermal transfer model for heterogeneous structure samples and mechanically bonded samples, showing the thickness of each layer and its related thermal properties. The key thermal properties of each layer (thermal conductivity, volumetric heat capacity, and interface thermal conductivity) are labeled as input parameters or fitting targets in the PWA-TDTR analysis. (a) $Ga_2O_3$-on-SiC heterogeneous structure; (b) GaN-on-Si heterogeneous structure; (c) GaN-on-diamond bonded sample.

To interrogate these complex multilayer systems, we employ PWA-TDTR as a depth-resolved thermal probe. This approach enables quantitative extraction of both interfacial thermal conductance and layer-specific thermal properties, thereby providing direct insight into how structural and material factors control heat transport across buried interfaces.

*3.1 Structural Overview of Samples*

All three representative heterostructures were prepared through fabrication routes commonly used in device manufacturing, allowing their thermal behavior to actually reflect real engineering conditions. Figure 2 summarizes the structural schematics used for thermal modeling: $Ga_2O_3$-on-SiC, GaN-on-Si, and GaN-on-diamond. Each sample exhibits a well-defined multilayer sequence spanning from hundreds of nanometers to tens of micrometers, and contains one or more buried interfaces that dominate cross-plane heat dissipation.

The $Ga_2O_3$-on-SiC sample was grown by metal–organic chemical vapor deposition (MOCVD) [30]. This configuration represents a prototypical ultra-wide-bandgap heterostructure, where $Ga_2O_3$ is integrated onto a high-thermal-conductivity SiC substrate. Despite the simplicity of its layer sequence (Figure 2a), the large acoustic and lattice mismatch at the $Ga_2O_3$/SiC interface introduces a significant thermal barrier, making this structure an ideal model system for evaluating buried interfacial resistance.

The GaN-on-Si sample also grown by MOCVD includes an AlN buffer layer and an AlGaN transition layer on the Si substrate. Such transition layers are indispensable for strain relaxation but simultaneously introduce additional buried interfaces and defect scattering centers that reduce effective heat transfer. For thermal modeling, these interlayers—with similar volumetric heat capacities [31,32]—along with their interfacial resistances were grouped into an equivalent transition layer (Figure 2b), allowing simplified yet physically consistent analysis of the thermal transport path.

The GaN-on-diamond heterostructure is particularly representative of state-of-the-art thermal management strategies for high-power GaN devices (Figure 2c). Diamond, with an intrinsic thermal conductivity approaching 1800 W/(m·K), provides an exceptional heat-spreading substrate compared with GaN [33]. However, the extreme lattice mismatch between GaN and diamond precludes direct epitaxy, necessitating mechanical or surface-activated bonding schemes [34]. These bonding processes often leave residual interlayers or voids that act as strong thermal bottlenecks, making the buried GaN–diamond interface the most critical junction to quantify.

The GaN-on-diamond sample studied here consists of a three-layer mechanically bonded stack: a ~1.8 μm GaN epilayer, a thinned Si interlayer (~22 μm), and a bulk diamond substrate. Heat generated in the GaN must cross multiple buried interfaces (Al/Si, Si/GaN, and GaN/diamond), each adding appreciable resistance along the vertical heat path. To fabricate this structure, a modified metal diffusion-assisted bonding process was employed: 5 nm Ti, a 5 nm Ti/Ag gradient layer, and 5 nm Ag were sequentially deposited on both GaN/Si and diamond surfaces by RF magnetron sputtering. The wafers were then pressed together at room temperature under 400 N axial force for 30 min, achieving hetero-integration without high-temperature annealing. Post-bonding ultrasonic inspection confirmed a bonding efficiency of 95.7%, validating the structural integrity of the buried interface.

Finally, a ~100 nm Al film was deposited by magnetron sputtering onto the surface of each sample, serving as the thermoreflectance transducer layer required for PWA-TDTR measurements [15].

*3.2 Experimental Design and Measurement Parameters*

All PWA-TDTR experiments were performed at room temperature. Modulation frequencies were selected according to the layer thicknesses and depths of buried interfaces, ensuring that the associated thermal penetration depths emphasized either near-surface or deeply buried regions. Representative dual-frequency combinations were used—53.8/203.8 kHz for ε-$Ga_2O_3$/SiC; 94/250 kHz for GaN/Si; and 23.5 kHz/10.6 MHz for GaN/diamond. Only the 10.6 MHz measurement corresponds to a conventional TDTR configuration, while all other frequencies used the low-frequency PWA-TDTR method.

A dual-frequency joint-fitting strategy was applied to each sample, leveraging the complementary sensitivity of high and low frequencies: higher frequencies emphasize surface-proximal properties, while lower frequencies provide deeper thermal-wave penetration. This complementary pairing significantly improves parameter identifiability.

Experimental boundary conditions were independently calibrated. The pump–probe spot size was determined by the beam-offset method and incorporated into the thermal model. The thickness of the Al transducer was determined by picosecond acoustics (121 nm for Ga$_2$O$_3$; ~99 nm for both GaN samples). The thermal conductivity of Al was obtained from its measured electrical resistivity using the Wiedemann–Franz relation [35] and validated against an Al/Si reference sample. The resulting thermal conductivity (~110 W/(m·K)) was fixed for all analyses. The volumetric heat capacities of all materials were adopted from an established literature database [36].

For all measured samples, most layer thicknesses were independently verified using picosecond acoustics or cross-sectional SEM. An exception is the Si interlayer in the GaN/diamond stack: the Si film is too thick for picosecond acoustic, and destructive cross-sectional SEM is not feasible due to the hardness of the diamond substrate. Consequently, we treat the Si thickness as a fitting parameter while keeping its volumetric heat capacity fixed to literature values. This approach is justified by the fact that the transient thermal response of a film is fundamentally governed by three combined parameters—$\sqrt{k_z C}$, $hC$, and $k_r/C$ [25,37]. When the heat capacity $C$ is known, the remaining parameters $k_r$, $k_z$, and $h$ can be uniquely determined through the fitting process.

To determine appropriate frequency pairs, we estimate the characteristic thermal penetration depth

$$\delta(f) \approx \sqrt{\frac{k}{\pi C f}},$$

using effective cross-plane thermal properties to gauge the depth probed at each frequency. Candidate $(f_{\text{low}}, f_{\text{high}})$ pairs were then evaluated by PWA-TDTR forward simulations to ensure that the two frequencies provide complementary, non-redundant sensitivity to the target parameters.

Because the three heterostructures differ substantially in thickness and thermal conductivity, the optimal frequency windows vary across samples. For ε-$Ga_2O_3$/SiC, (53.8 kHz, 203.8 kHz) yields penetration depths of several micrometers, adequately probing both the $Ga_2O_3$ layer and the $Ga_2O_3$/SiC interface. For GaN/Si, (94 kHz, 250 kHz) enables the lower frequency to sample the GaN, transition region, and shallow Si, while the higher frequency emphasizes the GaN layer and its interface. For GaN/diamond, the combination of 10.6 MHz and 23.5 kHz allows near-surface characterization of the Al/Si stack at high frequency (penetration depth ≈1.6 μm) and access to the GaN/diamond interface at low frequency (penetration depth ≈33.7 μm). The PWA waveform analysis further enhances sensitivity to the GaN/diamond interface while suppressing the influence of the deep diamond substrate. Quantitative penetration-depth estimates and representative sensitivity curves are provided in Supplementary Section S1 and Table S1.

## 4. Discussion: Physical Insights from Depth-Resolved PWA-TDTR

*4.1 Probing buried heat transport through periodic-waveform physics*

The periodic-waveform analysis (PWA) framework extends TDTR beyond conventional harmonic heating by reconstructing the full temporal waveform of surface temperature oscillations. In physical terms, this approach directly captures the thermal-wave dispersion governed by the heat equation in multilayer systems. Lower modulation frequencies correspond to longer thermal wavelengths and larger penetration depths, allowing the heat flux to interact with buried interfaces where phonon scattering and interlayer mismatch dominate. Thus, the modulation frequency in PWA-TDTR acts as a tunable filter for phonon transport length scales: high frequencies emphasize diffusive heat flow near the surface, whereas low frequencies reveal cumulative contributions from deeper, more weakly coupled interfaces.

This frequency-domain accessibility bridges the long-standing gap between surface-limited TDTR and steady-state thermoreflectance (SSTR), enabling a unified description of heat transport from nanometer to tens-of-micrometer depths within a single optical platform. The experimental results below illustrate how this tunable thermal-wave regime translates into a quantitative and physical understanding of interfacial phonon transport in three classes of heterostructures.

*4.2 $Ga_2O_3$-on-SiC: Phonon transmission across epitaxial mismatch*

The $Ga_2O_3$-on-SiC interface is an epitaxial, crystalline-crystalline junction with significant acoustic mismatch. To probe its buried thermal properties, PWA-TDTR measurements were conducted using two modulation frequencies, 53.8 kHz and 203.8 kHz, which selectively emphasize shallow and deep thermal paths. The laser spot size was calibrated to $w_0 = 11$ μm via the beam-offset method. The thermal model includes twelve parameters: the thermal conductivity, volumetric heat capacity, and thickness of both the Al transducer and $Ga_2O_3$ film; the anisotropic thermal conductivities of the 4H-SiC substrate ($k_{r,SiC}$ and $k_{z,SiC}$) together with its heat capacity; the interfacial thermal conductances at the Al/$Ga_2O_3$ ($G_1$) and $Ga_2O_3$/SiC ($G_2$) boundaries; and the optical spot size. This parameter set captures the essential physical quantities governing heat flow from the surface transducer into the deeply buried $Ga_2O_3$/SiC interface.

The Al transducer has a thermal conductivity of 110 W/(m·K), a thickness of 121±6 nm (picosecond acoustics), and a heat capacity from literature [36]. The $Ga_2O_3$ layer is 320±16 nm thick (SEM), with calibrated thermal properties of $k_{GaO}$ = 3.21±0.11 W/(m·K) and $C_{GaO}$ = 3.01±0.14 MJ/(m$^3$·K) obtained via negative-delay TDTR [38]. The Al/$Ga_2O_3$ interface conductance was previously benchmarked at 70±12 MW/(m$^2$·K). These quantities are fixed in the model to avoid strong correlations with substrate properties. The SiC heat capacity (2.1±0.06 MJ/(m$^3$·K)) is taken from literature [39], while its anisotropic thermal conductivities and the Ga$_2$O$_3$/SiC interfacial conductance ($G_2$) are treated as fitting parameters.

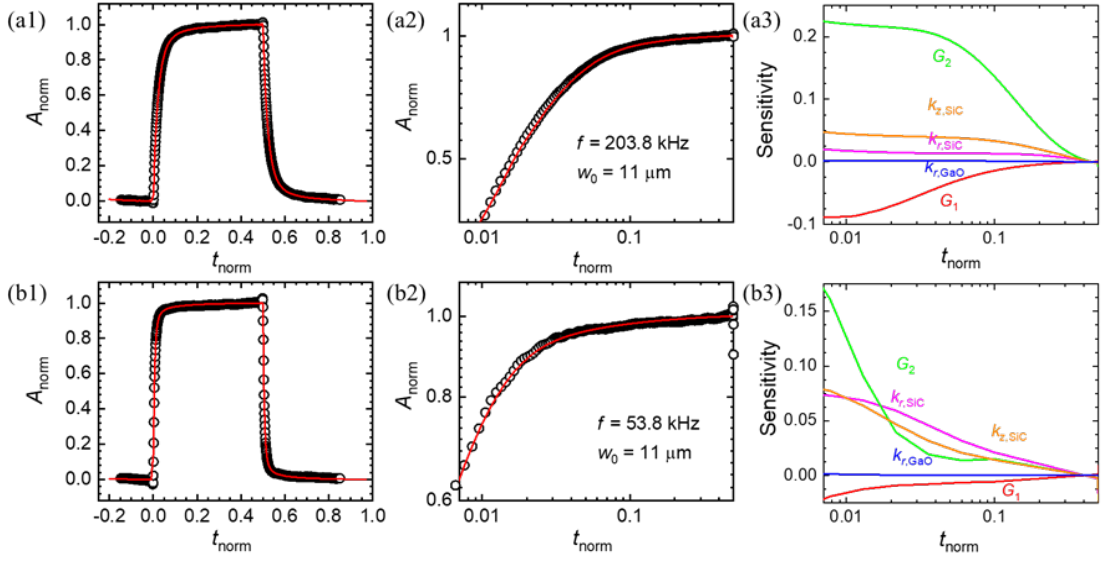

**Fig. 3.** PWA-TDTR measurement results and sensitivity analysis of the Ga$_2$O$_3$-on-SiC sample under different modulation frequencies. (a1-a3) show the full-period signal, the zoomed-in heating stage signal, and the corresponding sensitivity curves at a modulation frequency of 203.8 kHz. (b1-b3) present the same set of data at a modulation frequency of 53.8 kHz. The two datasets are jointly used for parameter fitting to enhance model identifiability and improve the accuracy of interfacial thermal conductance extraction.

Figure 3 presents the normalized thermoreflectance signals at both frequencies and their best-fit curves, along with sensitivity curves. At 203.8 kHz, the signals are dominated by $G_1$ and $G_2$, with $G_1$ exhibiting negative sensitivity, while $k_{r,\text{SiC}}$ and $k_{z,\text{SiC}}$ contribute weakly. At 53.8 kHz, sensitivity to both $k_{z,\text{SiC}}$ and $k_{r,\text{SiC}}$ strongly enhanced, with distinct trends from the higher frequency case. Sensitivity to $G_2$ also changes markedly, dropping steeply before leveling off, while $k_{r,\text{GaO}}$ remains insensitive. This frequency-dependent behavior enables effective decoupling of thermal responses from different depths.

Joint fitting of the dual-frequency data yields quantitative values for the buried interface and substrate: $G_2$ = 25±1.9 MW/(m$^2$·K) (7.6% uncertainty), $k_{z,\text{SiC}}$ = 280±55 W/(m·K) (19.5%), and $k_{r,\text{SiC}}$ = 400±58 W/(m·K) (14.6%). The extracted anisotropic conductivities of SiC are slightly lower than those reported by Qian et al. [40], likely reflecting differences in crystalline quality, but consistent with laser flash measurements by Wei et al. [41]. These results demonstrate that PWA-TDTR can simultaneously quantify buried interfacial conductance and anisotropic substrate transport in an intact heterostructure.

For completeness, additional FDTR measurements were conducted on the same sample using the same unknown parameters ($G_2$, $k_{z,\mathrm{SiC}}$, $k_{r,\mathrm{SiC}}$). As detailed in Supplementary Section S3, the FDTR results qualitatively agree with the PWA-TDTR analysis but exhibit larger uncertainties due to low-frequency noise and reduced sensitivity to SiC conductivities at high frequencies. This comparison highlights the advantage of the PWA-TDTR approach for resolving buried interfaces and anisotropic substrate properties.

Overall, the buried $Ga_2O_3$/SiC interface represents a substantial thermal bottleneck, emphasizing the importance of interface engineering for efficient heat dissipation in next-generation ultra-wide-bandgap semiconductor devices.

*4.3 GaN-on-Si: Transition-layer dominated heat redistribution*

The GaN-on-Si heterostructure introduces a more complex thermal transport pathway than $Ga_2O_3$-on-SiC, largely due to the buffer and transition layers required to mitigate lattice mismatch between GaN and Si. In this work, for thermal modeling purposes, the combined AlN and AlGaN buffer layers (0.22 μm and 0.46 μm, respectively) are represented as a single equivalent "transition layer" of 680 nm. This simplification is justified because their individual thicknesses and similar volumetric heat capacities fall below the resolution of the sensitivity analysis (Figure 2(b)). The consolidated layer effectively captures the dual role of this region as both a structural accommodation layer and a significant thermal barrier.

Experimentally, the sub-micron total thickness of these buffer layers is more than an order of magnitude smaller than both the optical spot size and the thermal diffusion length in our PWA-TDTR measurements. Consequently, the PWA-TDTR signals are sensitive primarily to the total cross-plane thermal resistance of this region, not to the detailed depth profile of its thermal conductivity. To avoid over-parameterization, we therefore model the AlN+AlGaN stack as a single isotropic "transition layer" characterized by an effective cross-plane thermal conductivity, $k_{z,\text{trans}}$. Reported cross-plane conductivities for similar heteroepitaxial films—10–30 W/(m·K) [42–44] for AlN on Si or SiO$_2$, and ≈ 7–10 W/(m·K) [45–47] for AlGaN and AlN/AlGaN digital alloys—place our fitted $k_{z,\text{trans}}$ within a physically reasonable range, consistent with the strong phonon scattering expected in such a defective, compositionally graded buffer. This effective-layer approach is standard in thermoreflectance studies of GaN heterostructures and is sufficient to capture the thermal impact of the transition region [48–51].

Probing these buried layers requires an extended thermal penetration depth. While conventional TDTR is largely insensitive to the underlying transition layer and Si substrate beneath the ~1 μm GaN film, PWA-TDTR extends modulation frequencies into the sub-100 kHz regime. We therefore employed dual-frequency measurements at 94.0 kHz and 250.0 kHz to disentangle the thermal responses of the GaN film, transition layer, and Si substrate.

Sensitivity analysis confirms distinct depth-dependent probing. At the higher frequency (250.0 kHz, Figure 4(a3)), the signal is dominated by the cross-plane conductivity of the transition layer and Si substrate, reflecting heat flux confined near the surface. At the lower frequency (94.0 kHz, Figure 4(b3)), sensitivity shifts deeper, enhancing response to both in-plane and cross-plane conduction in the Si substrate and even altering the sign of sensitivity to the transition layer. This frequency-dependent crossover allows the contributions of shallow and buried layers to be separated, enabling the simultaneous extraction of multiple thermal parameters.

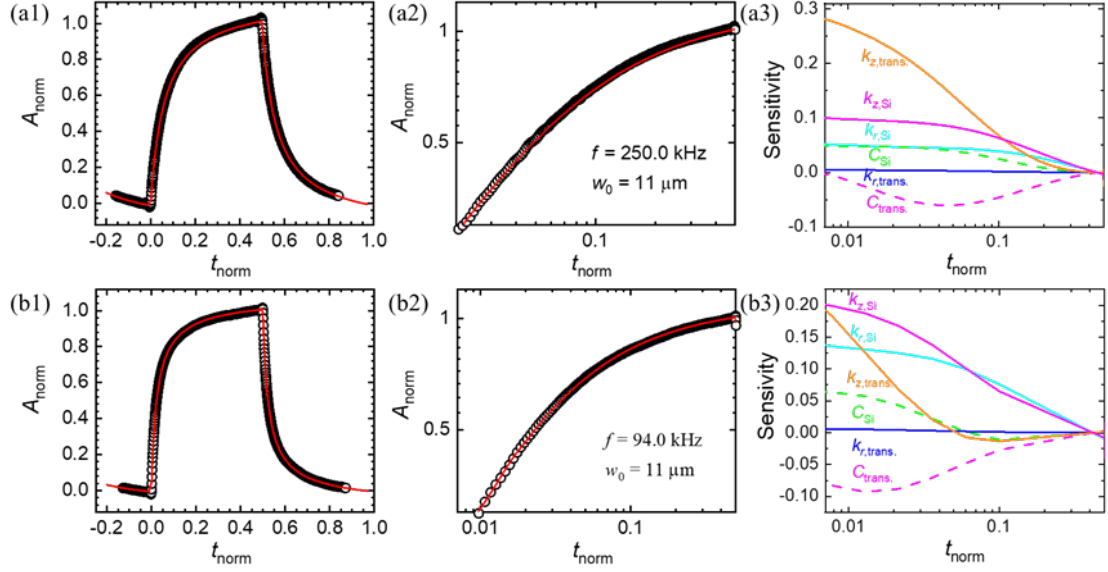

**Fig. 4.** PWA-TDTR measurement results and sensitivity analysis of the GaN-on-Si sample under different modulation frequencies. (a1-a3) show the full-period signal, the zoomed-in heating stage signal, and the corresponding sensitivity curves at a modulation frequency of 250 kHz. (b1-b3) present the same set of data at a modulation frequency of 94 kHz. The two datasets are jointly used for parameter fitting to enhance model identifiability and improve the accuracy of interfacial thermal conductance extraction.

In our model, both the equivalent transition layer and the Si substrate are treated as isotropic media ($k_r = k_z = k$), consistent with bulk Si's isotropy and the highly defective buffer stack. The reported values for $k_{trans}$ and $k_{Si}$ thus represent conductivities governing heat flow in both directions, even though the sensitivity curves distinguish directional contributions.

From joint fitting, the transition layer has a thermal conductivity of $k_{trans}$ = 14±0.51 W/(m·K) (3.6% uncertainty) and a volumetric heat capacity of $C_{trans}$ = 2.40±0.37 MJ/(m³·K) (15.3% uncertainty). These values align with a one-dimensional series-resistance effective medium approximation (EMA) for the constituent layers. Using layer thicknesses ($h_{AlN}$ = 0.22 μm and $h_{AlGaN}$ = 0.46 μm) and literature cross-plane thermal conductivities ($k_{z,AlN} \approx$ 18-25 W/(m·K) and $k_{z,AlGaN} \approx$ 10-14 W/(m·K)), the EMA predicts $k_{z,EMA} \approx$ 12-16 W/(m·K) and $C_{EMA} \approx$ 2.3-2.7 MJ/(m³·K). Out fitted results fall within these ranges, supporting their physical plausibility.

For the Si substrate, we obtain $k_{Si}$ = 133$\pm$6.9 W/(m·K) (5.2% uncertainty) and $C_{Si}$ = 1.61$\pm$0.11 MJ/(m$^3$·K) (7.0% uncertainty), in excellent agreement with bulk literature values [52,53]. The GaN layer, within measurement uncertainty, shows no pronounced anisotropy—likely due to competing scattering effects from dislocations (reducing in-plane transport) and thickness-induced boundary resistance (limiting cross-plane transport).

These results demonstrate the utility of PWA-TDTR for non-destructively probing buried thermal responses in epitaxial heterostructures. The technique directly retrieves coupled thermal properties of hidden layers without requiring destructive sample preparation. Physically, the low conductivity of the transition region confirms its role as a dominant thermal bottleneck in GaN-on-Si devices, where heat removal is already challenged by the relatively poor conductivity of the Si substrate. This underscores the critical need for buffer-layer engineering to manage thermal resistance in high-power GaN-on-Si electronics.

*4.4 GaN-on-Diamond: Limiting role of non-epitaxial bonding interfaces*

The GaN-on-diamond heterostructure presents one of the most challenging thermal configurations, characterized by a multilayer geometry and extreme thermal property contrasts across its interfaces. As illustrated in Figure 2(c), the simplified model consists of a GaN-on-Si structure mechanically bonded to a bulk diamond substrate via a Si interlayer. This architecture requires sequential heat transfer across multiple buried interfaces, making the overall thermal pathway highly sensitive to interfacial resistance. Physically, two interfacial bottlenecks are particularly critical: the Al/Si transducer interface ($G_1$) and the buried GaN/diamond interface ($G_2$), where phonon mismatch and imperfect bonding can severely impede heat flow.

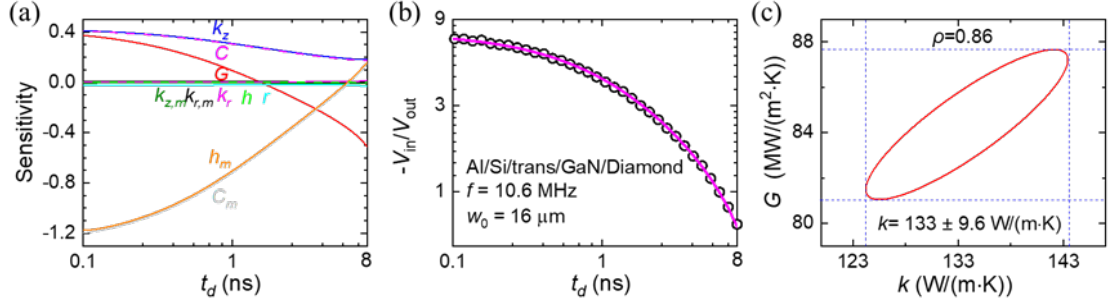

**Fig. 5.** TDTR measurement results of the GaN-on-Diamond sample: (a) Sensitivity analysis of this sample at a modulation frequency of 10.6 MHz; (b) TDTR experimental data; (c) 95% confidence region for the simultaneously extracted parameters $k_{Si}$ and $G_1$, along with the best-fit values, uncertainties, and correlation coefficient $\rho$.

To disentangle these contributions, a hybrid experimental strategy was employed. At high modulation frequency (10.6 MHz), conventional TDTR probes heat flow confined within the upper Al/Si region. Figure 5(b) shows the corresponding TDTR signal and model fit, while the associated sensitivity curves in Figure 5(a) and the joint confidence map in Figure 5(c) confirm that the signal is primarily sensitive to G1 and the thermal conductivity of the Si layer ($k_{Si}$). The best fits yield $G_1$=84.4 ± 3.3 MW/(m²·K) and $k_{Si}$ =133±9.6 W/(m·K). These values agree well with bulk references and results from the GaN-on-Si sample, indicating that thinning the Si layer does not significantly degrade its intrinsic thermal conductivity.

Accessing the deeply buried GaN/diamond interface requires extending thermal penetration depth, achieved using low-frequency PWA-TDTR at 23.5 kHz. At this frequency, the thermal wave traverses the combined ~24 μm Si+GaN stack. Since the final bonded Si interlayer thickness is not accessible non-destructively, we treated it as a fitting parameter, jointly inferred with $G_2$ to provide an in-situ estimate. Figure 6(a) presents the corresponding PWA-TDTR signal and fit, with the cooling segment highlighted in Figure 6(b). Sensitivity analysis in Figure 6(c) confirms a strong, coupled dependence on both the Si layer thickness ($h_{Si}$) and $G_2$. This enables their simultaneous extraction, yielding $h_{Si}$ = 22.0±0.46 μm and $G_2$ = 41±8.4 MW/(m²·K). Notably, the sensitivity analysis also shows that fixing the diamond thermal conductivity to its previously calibrated value (~1623 W/(m·K)) results in negligible changes to the fitted $h_{Si}$ and $G_2$ within the uncertainty range, confirming that the PWA-TDTR response is dominated by the Si interlayer and the interface, bot the bulk diamond.

The extracted $G_2$ value (41±8.4 MW/(m²·K)) lies in the mid-to-upper range for mechanically bonded GaN-on-diamond interfaces, suggesting partial phonon transmission despite the large acoustic mismatch and lack of epitaxial continuity. This intermediate conductance implies reasonably good interfacial conformity, likely aided by the diffusion-assisted interlayer used during bonding. It falls below values reported for high-quality bonding methods like surface-activated bonding or metal-interlayer epitaxy (80–120 MW/(m²·K)) [54,55] but exceeds those typical of untreated mechanical contacts (20–50 MW/(m²·K)) [56,57].

The principal measurement uncertainty stems from the extreme thermal diffusivity contrast between GaN and diamond. The rapid lateral heat spreading in diamond reduces the signal's sensitivity to the GaN/diamond boundary, thereby amplifying the effect of any microscopic interfacial imperfections (e.g., voids, roughness, or contamination). Furthermore, the thermal model's assumption of idealized interfacial contact may underestimate the influence of localized non-uniformities.

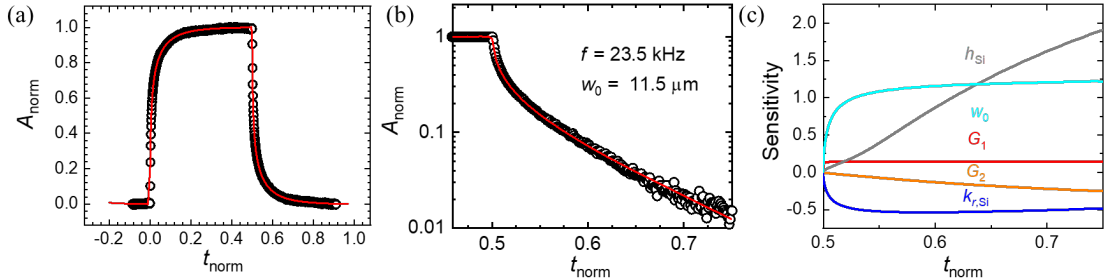

**Fig. 6.** PWA-TDTR measurement results of the GaN-on-Diamond sample: (a) The best fitting of the PWA-TDTR signals over the entire cycle and the thermal model; (b) Detailed presentation of the signal in the descending phase and the model fitting situation; (c) Sensitivity analysis corresponding to the descending phase

In summary, these results highlight the unique capability of low-frequency PWA-TDTR to probe deeply buried interfaces inaccessible to conventional TDTR. By jointly and in situ resolving both the Si interlayer thickness and the GaN/diamond interface conductance, this method avoids the need for destructive cross-sectional preparation. From a materials physics perspective, the measurement confirms that while diamond offers exceptional bulk thermal conductivity, the GaN/diamond interface often remains the limiting factor for heat dissipation. Consequently, advancing high-power GaN devices requires not only the integration of diamond substrates but also dedicated engineering of the buried bonding interface to minimize phonon scattering and thermal boundary resistance.

*4.5 Broader implications*

To provide a clear overview of the thermal properties obtained in this study, Table.1 summarizes the extracted thermal conductivities, volumetric heat capacities, and interfacial thermal conductances of the investigated heterostructures, together with representative literature values.

**Table.1.** The measured values of thermal conductivity ($k$), volumetric heat capacity ($C$), interfacial thermal conductance ($G$), and other properties of different samples using combined PWA-TDTR or TDTR techniques and comparison with literature values.

| Parameters [units] | Measured (Error) | Literature | Method | Sample |
|---|---|---|---|---|
| $k_{z,\text{SiC}}$ [W/(m·K)] | 280 (19.5%) | 280 [40] | PWA | Ga$_2$O$_3$-on-SiC |
| $k_{r,\text{SiC}}$ [W/(m·K)] | 400 (14.6%) | 347 [40] | PWA | |
| $G_{\text{GaO/SiC}}$ [MW/(m²·K)] | 25 (7.6%) | 20[a] [17] | PWA | |
| $G_{\text{GaN/Si}}$ [MW/(m²·K)] | 20.6 (3.6%) | — | PWA | GaN-on-Si |
| $k_{\text{Si}}$ [W/(m·K)] | 133 (5.2%) | 142 [52] | PWA | |
| $C_{\text{Si}}$ [MJ/(m³·K)] | 1.61 (7.0%) | 1.665 [53] | PWA | |
| $G_{\text{Al/Si}}$ [MW/(m²·K)] | 84.4 (3.9%) | 116 [13] | TDTR | GaN-on-diamond |
| $k_{\text{Si}}$ [W/(m·K)] | 133 (7.2%) | 142 [52] | TDTR | |
| $h_{\text{Si}}$ [μm] | 22.0 (2.1%) | — | PWA | |
| $G_{\text{GaN/diamond}}$ [MW/(m²·K)] | 41 (20.5%) | 20~50[b] [55] | PWA | |

Here, [a] represents the weakly bonded area, and [b] represents untreated mechanical contacts. For the GaN-on-Si sample, $k_{\text{Si}}$ denotes the isotropic thermal conductivity of the Si substrate ($k_r = k_z$).

Overall, the measured thermal conductivities and heat capacities agree well with established literature data. This alignment reflects the intrinsic material characteristics specific to the fabrication routes used in this study. For example, the thermal conductivity of $Ga_2O_3$ and the anisotropic properties of 4H-SiC are consistent with known trends, validating the PWA-TDTR method for probing both isotropic and anisotropic materials. Similarly, the effective thermal conductivity of the AlGaN/AlN transition layer agrees with a simple series-resistance average of its constituents, confirming the validity of the equivalent-layer modeling approach.

The extracted interfacial thermal conductances further highlight how growth and bonding methods govern phonon transport across interfaces. Epitaxial interfaces, such as Al/GaN, exhibit relatively high conductance (>100 MW/(m²·K)), while the mechanically bonded GaN/diamond interface shows a moderate value (~41 MW/(m²·K)). This places it between the ranges reported for simple mechanical contacts and advanced surface-activated bonding, underscoring the critical role of interfacial engineering in high-power device thermal management. By systematically benchmarking against literature, these results validate the robustness of PWA-TDTR and provide quantitative guidance for material selection and interface optimization in next-generation devices.

From a microscopic perspective, the measured TBC trends are consistent with contemporary phonon-transport models. For the epitaxial $Ga_2O_3$/SiC junction, the large acoustic mismatch and crystalline continuity suggest elastic phonon transmission as the baseline, with additional inelastic channels activated by anharmonicity and interfacial disorder [58–61]. Recent simulations indicate that defect-enriched transition regions near the interface can further suppress heat flow by shortening phonon lifetimes and redistributing mode populations [62].

For the GaN-on-Si system, the AlN/AlGaN transition region acts as a graded phonon-impedance layer. Atomistic calculations show that such buffers can create a vibrational density-of-states (VDOS) "bridge", partially aligning the spectra of the two materials and thereby enhancing phonon transmission compared to an abrupt interface [63].

Conversely, in mechanically bonded GaN-on-diamond structures, the absence of epitaxy combined with nanoscale roughness or voids promotes strong diffuse interfacial scattering. This can dramatically reduce TBC, even when the diamond substrate possesses ultrahigh bulk conductivity [64]. These microscopic mechanisms collectively explain why the experimentally extracted TBC values in this work lie between the weakly and strongly bonded limits. They underscore that interfacial chemistry, roughness, and buffer-layer design are as critical as bulk thermal properties for engineering heat flow in wide- and ultra-wide bandgap heterostructures.

## 5. Conclusions

This work demonstrates a frequency-tunable, Pulsed-Wideband Time-Domain Thermoreflectance (PWA-TDTR) framework to perform depth-resolved thermal metrology on three multilayer heterostructures: epitaxial $Ga_2O_3$/SiC and GaN/Si, and a mechanically bonded GaN/diamond stack. By leveraging modulation frequencies from 10 kHz to 10 MHz with joint multi-parameter sensitivity analysis, the method decouples heat-flow channels at different depths without requiring destructive sample preparation. High-frequency measurements resolve near-surface interfaces (e.g., Al/GaN) with high precision, while low-frequency operation extends thermal penetration to tens of micrometers, providing access to deeply buried interfaces (e.g., GaN/diamond) and revealing anisotropic transport in substrates like SiC.

Beyond metrology, the results deliver key physical insights into heat dissipation in device-relevant architectures. Interfacial thermal conductance is shown to depend critically on both bonding schemes and the presence of intermediate layers. Transition regions—even when sub-micrometer in thickness—can re-route heat and dominate the total thermal resistance. For the GaN/diamond structure, the measured thermal boundary conductance falls in the mid-to-upper range for mechanical bonding, illustrating that diamond's ultrahigh bulk conductivity does not guarantee system-level performance without concurrent engineering of phonon coupling at the buried interface. These findings yield direct design principles: interface quality (roughness, cleanliness, conformity), interlayer selection, and precise thickness control must be prioritized to mitigate thermal bottlenecks.

Overall, PWA-TDTR delivers "deep accessibility with layered resolvability" for complex heterostructures, establishing a reliable, nondestructive platform for screening and optimizing buried nonmetal–nonmetal interfaces essential to next-generation wide- and ultra-wide bandgap electronics. The method is readily extended to temperature-dependent and operando studies, where integration with microstructural characterization can further link interfacial chemistry and morphology to phonon transport.

**Credit author statement**

R.Y., and P.J. designed the research. M.Z. conceived the experiments, characterization, and modeling. M.Z., P.J., and R.Y. wrote the manuscript. All authors participated in the discussion of the research.

**Declaration of competing interest**

The authors declare that they have no known competing financial interests or personal relationships that could have appeared to influence the work reported in this paper.

**Acknowledgements**

The authors extend their sincere gratitude to Mr. Tao Chen of HUST for performing the FDTR measurements. This work was supported by the National Key Research and Development Program of China (Grant No. 2022YFB3803900), the National Natural Science Foundation of China (NSFC) (Grant No. T2588301), and the National Natural Science Foundation of China (NSFC) (Grant No. 52376058).